\documentclass[aps, prb, reprint, groupedaddress, superscriptaddress, showpacs]{revtex4-1}

\usepackage{graphicx}
\graphicspath{{./figs/}}
\usepackage{commath, amsmath, amssymb, amsfonts}
\usepackage{hyperref}
\usepackage{physics, siunitx}
\usepackage{epstopdf}

\usepackage[normalem]{ulem}
\usepackage[usenames,dvipsnames]{xcolor}

\renewcommand{\va}[1]{\mathbf{#1}}
\newcommand{\scase}[1]{\mathcal{{#1}}}
\newcommand{\field}[1]{\boldsymbol{\mathcal{{#1}}}}
\usepackage{array}
\newcolumntype{P}[1]{>{\centering\arraybackslash}p{#1}}
\newcolumntype{M}[1]{>{\centering\arraybackslash}m{#1}}
\allowdisplaybreaks
\newcommand{\ie}{i.e.~}

\newcommand{\eg}{e.g.~}

\hypersetup{
	colorlinks   = true, 
	urlcolor     = blue, 
	linkcolor    = blue, 
	citecolor   = blue 
}

\begin{document}


\title{Linear and nonlinear optical response of crystals using length and velocity gauges: Effect of basis truncation}


\author{Alireza Taghizadeh}
\email{ata@nano.aau.dk}
\author{F. Hipolito}
\affiliation{Department of Physics and Nanotechnology, Aalborg University, DK-9220 Aalborg {\O}st, Denmark}

\author{T. G. Pedersen}
\affiliation{Department of Physics and Nanotechnology, Aalborg University, DK-9220 Aalborg {\O}st, Denmark}
\affiliation{Center for Nanostructured Graphene (CNG), DK-9220 Aalborg {\O}st, Denmark}
\date{(Received XX XXXXXX 2017, published XX XXXXXX 2017}%


\date{\today}

\begin{abstract}
We study the effects of a truncated band structure on the linear and nonlinear optical response of crystals using four methods. These are constructed by (i) choosing either length or velocity gauge for the perturbation and (ii) computing the current density either directly or via the time-derivative of the polarization density. In the infinite band limit, the results of all four methods are identical, but basis truncation breaks their equivalence. In particular, certain response functions vanish identically and unphysical low-frequency divergences are observed for few-band models in the velocity gauge. Using hexagonal boron nitride (hBN) monolayer as a case study, we analyze the problems associated with all methods and identify the optimal one. Our results show that the length gauge calculations provide the fastest convergence rates as well as the most accurate spectra for any basis size and, moreover, that low-frequency divergences are eliminated.
\end{abstract}


\maketitle

\section{Introduction}


The optical response of crystals is essential for countless technological applications of solids as well as for characterization of materials. In semiconductors and insulators, optics provide access to important features of the band structure including band gaps and transitions at high-symmetry points \cite{Basu2002, Yu2010}.
The optical response can be characterized by the linear response as well as various nonlinear responses, \eg second/third harmonic generation (SHG/THG), optical rectification (OR), sum/difference-frequency generation, etc\cite{Boyd2008}. Several nonlinear optical (NLO) phenomena have important scientific and technical applications at energies ranging from the THz to visible wavelengths such as in laser technology, optical communication, bio-molecular detection, and surface characterization \cite{Shen2002, Boyd2008}. The interest in NLO processes has recently grown dramatically due to the large response and exotic phenomena observed in two-dimensional (2D) materials such as graphene \cite{Hendry2010, Hong2013, Wang2016, Mikhailov2016, Semnani2016, Sun2016}, hexagonal boron nitride (hBN) \cite{Li2013, Pedersen2015, Brun2015, Hipolito2016}, and transition metal dichalcogenides \cite{Kumar2013, Wang2013, Trolle2014, Mak2014, Mak2016}. 

From a theoretical point of view, a reliable method for the computation of linear and nonlinear optical response functions based on the material band structure is crucial. For the linear optical response, such calculations are now routinely performed and excellent agreement with experiments is obtained, see \eg Ref. \onlinecite{Yu2010}. However, the calculation of the NLO response of crystals remains an open subject as various methods for calculation, \eg the choice between length and velocity gauges, frequently generate different results \cite{Kobe1979, Moss1990, Sipe1993, Aversa1995, Sipe2000, Madsen2002, Rzazewski2004, Virk2007, Chen2009, Dong2014, Wu2015, Foldi2017, Ventura2017, Wang2017}.
An extreme example of these differences emerges when considering systems whose electronic properties can be captured accurately by a two-band Hamiltonian, \eg hBN \cite{Ribeiro2011, Pedersen2015} or the low energy properties of biased bilayer graphene \cite{Castro2007, Castro2009, Brun2015}. 
In such systems, the evaluation of the second-order (in fact, all even-order) response in the velocity gauge is identically zero at all frequencies, whereas the equivalent calculation using the length gauge leads to finite results \cite{Pedersen2015, Hipolito2016}. Moreover, computing the NLO response in the length gauge is significantly more complex than the velocity gauge due to the appearance of generalized derivatives (GDs) \cite{Aversa1995, Sipe2000, Pedersen2015}. 
In spite of the above-mentioned, it can be shown that gauge invariance is upheld if a complete basis set is used for both calculations \cite{Rzazewski2004, Ventura2017}.  
However, for many practical reasons, both analytic and numerical approaches to the calculation of the optical response rely on truncated basis sets, that break the gauge invariance. 
The influence of basis truncation on the optical response in length and velocity gauges was discussed recently in Ref. \onlinecite{Ventura2017}, where it was predicted qualitatively that at low frequencies, the contributions from the omitted bands in the velocity gauge can be considerable. The gauge freedom, however, is not the only source of differences between commonly used computational approaches. Hence, different choices exist for the observable providing the optical response. Moreover, the GD in length gauge calculations can be evaluated by its definition but, also, circumvented using an approximate sum rule presented in Refs.~\onlinecite{Aversa1995, Rashkeev2001}. 
 
In this paper, all of these alternatives will be examined, emphasizing the effects of basis truncation for a real material (hBN) and in a broad frequency range. Hence, we compare four computational approaches to the optical response including the linear optical conductivity (OC), OR, SHG and THG of periodic systems and study the convergence of each approach as a function of basis size. 
The four methods consist of the combinations of two choices of gauge, \ie length and velocity gauges, and two ways of computing the current density response: direct evaluation of the current density or via the time-derivative of the polarization density. In addition, we investigate the effects arising from evaluating the GDs by the above-mentioned sum rule for a truncated basis set. We choose monolayer hBN as a test case due to the simplicity of the band structure and the pronounced two-band character of the material.
In order to have access to a variable-size basis set, we use an empirical pseudopotential Hamiltonian \cite{Cohen1966, Chelikowsky1976} that reproduces the low-energy properties of hBN monolayers \cite{Ribeiro2011, Pedersen2015}. 
We find that the length gauge approach generates the most accurate results among the considered methods for any basis size. Moreover, we study the effect of basis set truncation on the unphysical zero-frequency divergences plaguing velocity gauge calculations \cite{Moss1990, Sipe1993, Aversa1995}.
Thus, our results provide guidelines for choosing the optimal computational method for the optical response based on a truncated band structure. 
This is essential in cases, where the number of available bands is typically limited such as tight-binding (TB) models. Similarly, many-body calculations employing the Bethe-Salpeter equation frequently rely on a truncated band structure (see e.g. Ref.~\cite{Trolle2014}) due to the computational complexity. In both cases, an optimal combination of gauge and observable is crucial.

The paper is organized as follows. First, we present the pseudopotential approach for computing the electronic band structure of hBN using the energy dispersion of a two-band TB model for parameterization. 
Then, the dynamical equation of motion is reviewed in section~\ref{sec:OpticalResponse} and its perturbative solution is derived up to the third-order. Based on this solution, we numerically compute and compare the linear, SHG, OR and THG conductivity spectra for hBN monolayer using the four above-mentioned approaches, and analyze the influence of basis truncation on the calculated spectra. Finally, a summary of results is presented in section~\ref{sec:Conclusion}. 

\begin{figure}[b]
	\includegraphics[width=0.49\textwidth]{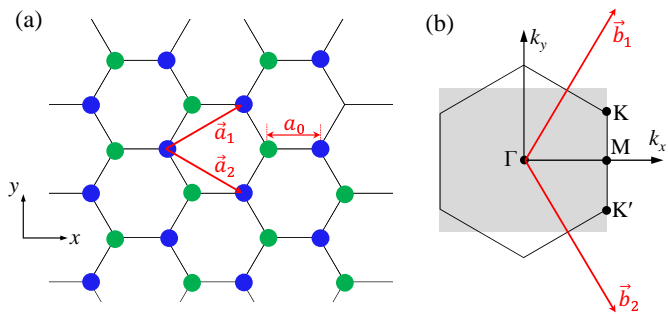}
	\caption{(a) Illustration of a 2D honeycomb lattice with a two-atom unit cell, \ie a boron (green) and a nitrogen (blue) atom, and the primitive vectors $\va{a}_1$ and $\va{a}_2$. 
		(b) The first Brillouin zone (BZ) with primitive reciprocal vectors $\va{b}_1$ and $\va{b}_2$. The shaded region shows the integration area used in this work, which is equivalent to the first BZ. }
	\label{fig:Schematic}
\end{figure}

\section{Pseudopotential Hamiltonian}
The hBN monolayer is a 2D crystal with a honeycomb lattice consisting of two different atoms per unit cell as depicted in Fig.~\ref{fig:Schematic}(a), with primitive lattice vectors $\va{a}_1=a(\sqrt{3}\va{e}_x+\va{e}_y)/2$ and $\va{a}_2=a(\sqrt{3}\va{e}_x-\va{e}_y)/2$, 
where $a$ is the lattice constant related to the inter-atomic distance $a_0$ via $a=\sqrt{3}a_0$ ($a_0=\SI{1.45}{\angstrom}$ for the hBN \cite{Ribeiro2011,Pedersen2015}). 
The first Brillouin zone (BZ), associated high-symmetric points and primitive reciprocal vectors $\va{b}_1$ and $\va{b}_2$ are illustrated in Fig.~\ref{fig:Schematic}(b). It should noted that, throughout the text, all vectors are indicated by bold letters, and Greek subscripts/superscripts denote the Cartesian components of vectors and tensors.

The characterization of the electronic properties of the system is based on an empirical pseudopotential Hamiltonian \cite{Phillips1958, Cohen1966, Chelikowsky1976}.
In this method, the real potential that governs the motion of electrons in the system is replaced by a simple effective potential, the \emph{pseudopotential}, which lumps together the effects of core electrons as well as crystal nuclei \cite{Yu2010}. 
By writing the Hamiltonian eigenstates as $\phi(\va{r})=\exp(i\va{k} \cdot \va{r})u(\va{r})$, with a crystal momentum $\va{k}$ and lattice-periodic function $u(\va{r})$, the Schr\"{o}dinger equation is transformed into an eigenvalue problem in reciprocal space as \cite{Cohen1966}
\begin{align}
	\label{eq:PsPotEigen} 
	\sum_{\va{G}'} \Big[ \frac{ \hbar^2 }{ 2 m } | \va{k} +\va{G} |^2 \delta_{\va{G}\va{G}'} +\mathcal{V}_{ \va{G}-\va{G}' } \Big] u_{\va{G}'} = E \, u_\va{G} \, ,
\end{align}
where $\va{G}$ and $\delta_{\va{G}\va{G}'}$ represent the reciprocal lattice vectors and Kronecker delta, respectively. The Fourier coefficients of the pseudopotential in the reciprocal space read $\mathcal{V}_{\va{G}}$, while $u_\va{G}$ and $E$ denote the eigenvectors (the Fourier coefficients of $u$) and eigenvalues, respectively. 
The Fourier coefficients of the pseudopotential $\mathcal{V}_{\va{G}}$ can be decomposed into symmetric and anti-symmetric parts, the so-called \emph{form factors} $\mathcal{V}_\va{G}^{S}$ and $\mathcal{V}_\va{G}^{AS}$ as \cite{Cohen1966}
\begin{align}
	\mathcal{V}_{\va{G}} \equiv \mathcal{V}_{\va{G}}^S \cos( \va{G} \cdot \boldsymbol{\tau} ) +i \mathcal{V}_{\va{G}}^{AS} \sin(\va{G} \cdot \boldsymbol{\tau}) \, ,
\end{align}
where $2\boldsymbol{\tau}=a/\sqrt{3}\va{e}_x$ is the vector connecting the two atoms in the unit cell. 
\begin{table}
	\caption{Pseudopotential form factors (in eV) for monolayer hBN obtained by fitting the band structure to a  TB model \cite{Ribeiro2011}. The reciprocal vectors $\va{G}$ are normalized by $2\pi/a$.} 
	\label{tb:Pseudopotential}
	\centering
	\begin{tabular}{cccc} 
		\hline \hline 
		$|\va{G}|^2$ & 4/3 & 4 & 16/3 \\
		\hline 
		$\mathcal{V}_{\va{G}}^{S}$ & 10.785 & 1.472 & 7.8 \\
		$\mathcal{V}_{\va{G}}^{AS}$ & 8.007 & 0 & 3.697 \\
		\hline \hline 
	\end{tabular}
\end{table} 

\begin{figure}[t]
	\includegraphics[width=0.49\textwidth]{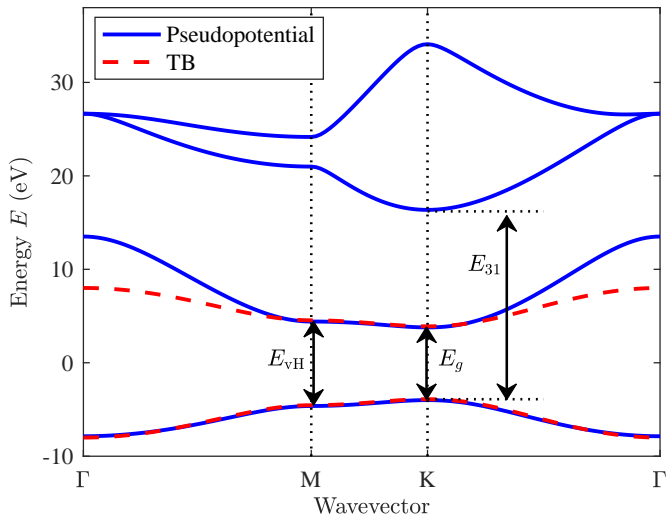}
	\caption[Bandstructure of hBN]{Band structures of hBN monolayer computed by the two $\pi$-band TB model (red) and pseudopotential Hamiltonian (blue). The TB parametrization follows Ref.~\onlinecite{Ribeiro2011} and the pseudopotential Hamiltonian is parametrized according to the form factors presented in Table~\ref{tb:Pseudopotential}, using a total of 43 reciprocal lattice vectors for the eigenvalue problem, Eq.~(\ref{eq:PsPotEigen}). Three important transition energies are defined: at the band gap $E_g \equiv E_2(\mathrm{K})-E_1(\mathrm{K})$, van Hove (vH) singularity $E_\mathrm{vH} \equiv E_2(\mathrm{M})-E_1(\mathrm{M})$, and $E_\mathrm{31} \equiv E_3(\mathrm{M})-E_1(\mathrm{M})$, where $E_i$ denote the energy of the $i$-th band.}
	\label{fig:BandStructure}
\end{figure}

We limit the calculation of the form factors to the first four smallest $|\va{G}|$ with squared magnitude of 0, 4/3, 4, 16/3 (normalized by $(2\pi/a)^2$).  
%
The value of $\mathcal{V}_0^{S}$ is not important, since it only shifts the energies, while the anti-symmetric form factors for $|\va{G}|^2 = 0$ and $4(2\pi/a)^2$ are not important since $\va{G}\cdot\boldsymbol{\tau}$ is 0 or $\pm\pi$, respectively. In addition, we adopt the spherical approximation for the pseudopotential \cite{Cohen1966}, which reduces the total number of unknown form factors to five: three symmetric and two anti-symmetric form factors.
The form factors used for hBN monolayer are presented in Table~\ref{tb:Pseudopotential}. These were determined by fitting the pseudopotential band structure to the low-energy part, \ie the vicinity of the $\mathrm{K}$ and $\mathrm{M}$ $k$-points, of the TB band structure for hBN monolayer \cite{Ribeiro2011, Margulis2013, Pedersen2015} employing a nearest-neighbor hopping integral of $\gamma_0 = 2.33$ eV and on-site energies of $\pm 3.9$ eV. 


In Fig.~\ref{fig:BandStructure}, we compare the energy dispersions of the pseudopotential and TB Hamiltonians.  
The pseudopotential Hamiltonian reproduces accurately the energy dispersion in the vicinity of the $\mathrm{K}$ and $\mathrm{M}$ points, but deviates from the TB model at the BZ center, \ie the $\mathrm{\Gamma}$ point, similarly to 
the ab-initio calculations of Ref.~\onlinecite{Ribeiro2011}.
In this manuscript, we have employed 43 reciprocal lattice vectors, which generates a total of 43 bands, 41 of which have dispersions above the TB conduction band. This large number of bands allows us to study the convergence of the optical response as a function of basis size, \ie the number of bands used in the calculation. The band gap and van Hove transition energies are $E_g=7.78$ eV and $E_\mathrm{vH}=9.04$ eV, respectively, while the transition to the second conduction band occurs at an energy of $E_{31}=20.4$ eV (see Fig. \ref{fig:BandStructure}).

\section{Current density response \label{sec:OpticalResponse}}
Here, we briefly review the calculation of the optical response of a periodic system in equilibrium under the influence of an external electromagnetic field. The periodic system is characterized by an unperturbed Hamiltonian, $\hat{H}_0$, and an external time-dependent perturbation $\hat{V}(t)$, such that the total Hamiltonian reads $\hat{H} = \hat{H}_0 + \hat{V}(t)$. 
The unperturbed Hamiltonian, $\hat{H}_0$, leads to the pseudopotential eigenvalue problem Eq.~(\ref{eq:PsPotEigen}). The form of the external potential depends on the gauge choice and in the length gauge $\hat{V}_l(t) = e \hat{\va{r}} \cdot \field{E}$ whereas, in the velocity gauge, $V_v(t)=e(\hat{\va{p}} \cdot \field{A} +e\field{A}^2/2 )/m$. Here, $\field{E}$ and $\field{A}$ are the electric field and vector potential, respectively.
Both choices have their merits and shortcomings in the context of periodic systems. The latter benefits from the fact that the matrix elements of the momentum operator are easily computed, but is plagued by spurious divergences 
since the electric field has to be mapped to the vector potential via $\field{E}=-\partial \field{A} / \partial t$. 
In contrast, the former requires a more elaborate calculation of the optical response, but circumvents the unphysical divergences at zero frequency \cite{Aversa1995}. Throughout this work, the external electromagnetic field is defined by its decomposition into harmonic components,
\begin{equation}
	\field{E}(t) = \dfrac{1}{2} \sum_p \sum_\alpha \va{e}_\alpha \mathcal{E}_\alpha(\omega_p) e^{-i\omega_p t} \, ,
\end{equation}
where the $p$-summation is performed over both positive and negative frequencies. 

The optical response calculation relies on the evaluation of the time-dependent density operator, $\hat{\rho}(t) = \sum_{mn} \rho_{mn}(t)\ket{m}\bra{n}$, governed by the quantum Liouville equation $i\hbar \partial \hat{\rho}(t)/\partial t = [\hat{H},\hat{\rho}(t)] $.
This equation is solved perturbatively (see Appendix~\ref{sec:AppendixA} for details) to obtain the optical response either by evaluating directly the expectation value of the current density operator, $\va{J}(t) = \tr\{\hat{\va{J}}\,\hat{\rho}(t) \}$, or by computing the time-derivative of the expectation value of the polarization density operator, $\va{J}(t) \equiv \partial \va{P}(t)/\partial t = \partial \tr\{ \hat{\va{P}} \, \hat{\rho}(t) \}/\partial t$ \cite{Sharma2004, Pedersen2015}. Here, the current and polarization density operators read $\hat{\va{J}} = -eg\hat{\va{p}}/(mA)$ and $\hat{\va{P}} = -eg\hat{\va{r}}/A$, respectively, where $g=2$ accounts for the spin degeneracy and $A$ is the crystal area.
The combination of two gauges and two ways of evaluating the current density response leads to a total of four approaches to compute the response as summarized and labeled in Table~\ref{tb:Gauges}. 
Below, we briefly discuss the important details regarding the calculation of linear and nonlinear current density response in these four approaches using the perturbative expansion of the density matrix.

\begin{table}[b]
	\caption[Methods to compute the current response]{
	\label{tb:Gauges}
	Four methods for computing the current density response and their respective labels. The methods arise from the combination of gauge freedom and choice of physical observable under consideration, \ie direct evaluation of the current or the time derivative of the polarization. Here, $\hat{\va{r}}$, $\hat{\va{p}}$, $\field{E}$, and $\field{A}$ represent the position operator, momentum operator, electric field, and vector potential, respectively.} 
	\centering
	\begin{tabular}{ccc}
		\hline \hline 
		label & $\hat{V}(t) \propto $ &  $\va{J}(t) \propto $ \\  
		\hline
		$\scase{A}$ & $ \hat{\va{r}} \vdot \field{E}$ & $\tr\{ \hat{\va{p}} \; \hat\rho(t) \} $ \\
		$\scase{B}$ & $ \hat{\va{r}} \vdot \field{E}$ & $ \partial \tr\{\hat{\mathbf{r}} \; \hat\rho(t) \} /\partial t$ \\
		$\scase{C}$ & $\hat{\va{p}} \vdot \field{A} + e \field{A}^2 /2$ & $\tr\{ ( \hat{\mathbf{p}} +e \field{A}/2 ) \; \hat\rho(t) \} $ \\ 
		$\scase{D}$ & $\hat{\va{p}} \vdot \field{A} + e \field{A}^2 /2$ & $\partial \tr\{\hat{\va{r}} \; \hat\rho(t) \} / \partial t$  \\
		\hline \hline 
	\end{tabular}
\end{table}

 
The evaluation of the density matrix elements, and direct calculation of the current density response in the velocity gauge is a rather straightforward problem, since the momentum operator is a well-defined operator in periodic systems.
%
In contrast, both the perturbation evaluation in the length gauge and the calculation of the polarization density involve the ill-defined (in periodic systems) position operator, $\hat{\va{r}}$.
In spite of the \textit{prima facie} problems associated with the position operator, it has been shown in Ref.~\onlinecite{Aversa1995}, and references therein, that the optical response can be computed in this gauge, by separating formally the interband and intraband parts of the position operator as $\va{\hat{r}}=\hat{\va{r}}^{(e)}+\hat{\va{r}}^{(i)}$
\begin{subequations}
	\label{eq:PositionOperator}
	\begin{align}
		&\va{r}_{nm}^{(e)} \equiv \mel{\phi_{n\va{k}}}{\hat{\va{r}}^{(e)}}{\phi_{m\va{k}'}} = (1-\delta_{nm}) \delta_{\va{k} \va{k}'} \va{\Omega}_{nm} \, , \\
		&\va{r}_{nm}^{(i)} \equiv \mel{\phi_{n\va{k}}}{\hat{\va{r}}^{(i)}}{\phi_{m\va{k}'}} = \delta_{nm} \big( \va{\Omega}_{nn} + i\gradient_{\va{k}} \big) \delta_{\va{k}\va{k}'} \, ,
	\end{align}
\end{subequations}
where the generalized Berry connections are defined as
\begin{equation}
	\va{\Omega}_{nm} \equiv \mel{u_{n\va{k}}}{i \gradient_{\va{k}}}{u_{m\va{k}}} = \dfrac{i}{A_\mathrm{uc}} \int_{\text{uc}} u_{n\va{k}}^* \gradient_{\va{k}} u_{m\va{k}} \dd[2]\va{r} \, ,
\end{equation}
with the cell-periodic functions $u_{n\va{k}}$ and unit cell area $A_\mathrm{uc}$. To simplify the notation, we frequently suppress the explicit dependence of quantities on wavevector. 
The interband matrix elements of position and momentum operators are related by $im\va{\Omega}_{nm}=\hbar \va{p}_{nm}/E_{nm}$ \cite{Gu2013} for $n\neq m$, where $E_{nm} \equiv E_n-E_m$. 
In addition, the intraband part of the position operator leads to appearance of the GD denoted typically by $()_{;\va{k}}$ \cite{Aversa1995, Sipe2000}. For any simple operator (diagonal in $\va{k}$), $\hat{O}$, the following expressions are then derived \cite{Aversa1995}
\begin{subequations}
	\label{eq:GenDerivative}
	\begin{align}
		&\mel{\phi_{n\va{k}}}{[\hat{\va{r}}^{(i)},\hat{O}]}{\phi_{m\va{k}'}} = i \delta_{\va{k}\va{k}'} \big(O_{nm} \big)_{;\va{k}} \, , \\
		&\big(O_{nm} \big)_{;\va{k}} \equiv \gradient_{\va{k}} O_{nm} - i[\va{\Omega}_{nn} - \va{\Omega}_{mm}] O_{nm} \, .
	\end{align}
\end{subequations}
In addition, by virtue of the canonical commutation relation, \ie $\comm{\hat{r}^\alpha}{\hat{p}^\beta}=i\hbar\delta_{\alpha\beta}$, and, by separating the interband/intraband parts of position operator, a sum rule is derived for the GD
\begin{equation}
	\label{eq:SumRule}
	\big(p_{mn}^\beta \big)_{;k^\alpha} = \hbar\delta_{\alpha\beta}\delta_{mn}+\dfrac{\hbar}{m} \sum_{l} \bigg[ \dfrac{\bar{\delta}_{ml} p_{ml}^\alpha p_{ln}^\beta}{E_{ml}} - \dfrac{\bar{\delta}_{ln} p_{ml}^\beta p_{ln}^\alpha}{E_{ln}} \bigg] \, ,
\end{equation}
where we introduce $\bar{\delta}_{ml} \equiv 1-\delta_{ml}$. Equivalent procedure was used for the GD evaluation in Refs. \onlinecite{Aversa1995, Sipe2000, Rashkeev2001}.
The basis truncation breaks this sum rule, thus opening a door for additional convergence problem as discussed in section~\ref{sec:Results}.
Making use of the perturbative solution of the density matrix, Eq.~(\ref{eq:DensityElements}), and the above-mentioned results, we evaluate the current density up to the third-order in the electric field strength.

We begin by addressing in detail the four possible methods to compute the linear response.
Without loss of generality, the first-order current density $J_\lambda^{(1)}(t)$ reads
\begin{equation}
	J_\lambda^{(1)}(t) = \sum_p \sum_\alpha \sigma_{\lambda\alpha}^{(1)}(\omega_p)\mathcal{E}_{\alpha}(\omega_p) e^{-i\omega_p t} \,.
\end{equation}
The different methods of calculation lead to four conductivity tensors $\sigma_{\lambda\alpha}^{(1)}$, defined as
\begin{subequations}
	\label{eq:FirstOrder}
	\begin{align}
	\label{eq:FirstOrderA} \sigma_{\lambda\alpha}^{\scase{A}(1)} &\equiv C_e \sum_{\substack{\va{k},n,m \\ n \neq m}} \dfrac{p^\lambda_{nm} g_{mn}^\alpha}{E_{mn}} - C_i \sum_{\va{k},n} \frac{p^\lambda_{nn}}{\hbar\omega_p} \pdv{f_{n}}{k^\alpha} \, , \\
	\label{eq:FirstOrderB} \sigma_{\lambda\alpha}^{\scase{B}(1)} &\equiv C_e \sum_{\substack{\va{k},n,m \\ n \neq m}} \dfrac{p^\lambda_{nm} g_{mn}^\alpha \hbar\omega_p}{E_{mn}^2} - C_i \sum_{\va{k},n} \frac{m}{\hbar^2\omega_p} 
	\pdv{E_{n}}{k^\lambda} \pdv{f_{n}}{k^\alpha} \, , \\
	\label{eq:FirstOrderC} \sigma_{\lambda\alpha}^{\scase{C}(1)} &\equiv C_e \sum_{\va{k},n,m} \dfrac{p^\lambda_{nm} g_{mn}^\alpha}{\hbar\omega_p} + \frac{ie^2}{m A_\mathrm{uc} \omega_p}\delta_{\lambda\alpha} \, , \\
	\label{eq:FirstOrderD} \sigma_{\lambda\alpha}^{\scase{D}(1)} & \equiv \sigma_{\alpha\lambda}^{\scase{A}(1)} \, .
	\end{align}
\end{subequations}
Here, $f_{nm} \equiv f_n-f_m$ with $f_n \equiv f(E_n)$ the Fermi-Dirac distribution and the summation over $\va{k}$ implies an integral over the BZ, \ie $(2\pi)^D \sum_\va{k} \rightarrow A \int_{\mathrm{BZ}} \dd[D]{\va{k}}$ with $D$ the dimension ($D$ = 2 for hBN). Also, the indices $m,n,l\in \{1,2,\cdots\}$ run over all the bands, and the constant coefficients $C_e$ and $C_i$ and variable $g_{mn}^\alpha$ are defined as
\begin{equation}
	C_e \equiv C_i \dfrac{\hbar}{ m } \equiv \dfrac{i g e ^2 \hbar}{2 m ^2 A}  \enskip,\enskip g_{mn}^\alpha \equiv \dfrac{f_{nm} p^{\alpha}_{mn}}{\hbar\omega_p-E_{mn}} \, . 
	\label{eq:FirstConstants}
\end{equation}
Similarly, the second- and third-order current density responses are determined by
\begin{align}
	J_\lambda^{(2)}(t) = \sum_{p,q} &\sum_{\alpha,\beta} \sigma_{\lambda\alpha\beta}^{(2)}(\omega_p,\omega_q)\mathcal{E}_{\alpha}(\omega_p)\mathcal{E}_{\beta}(\omega_q) e^{-i(\omega_p+\omega_q) t} \, , \\
	J_\lambda^{(3)}(t) = \sum_{p,q,s} &\sum_{\alpha\beta\gamma} \sigma_{\lambda\alpha\beta\gamma}^{(3)}(\omega_p,\omega_q,\omega_s) \nonumber \\
	&\mathcal{E}_\alpha(\omega_p) \mathcal{E}_\beta(\omega_q) \mathcal{E}_\gamma(\omega_s) e^{-i(\omega_p+\omega_q+\omega_s)} \, . 
\end{align} 
Given their complicated form, the expressions for the quadratic and cubic conductivity tensors can be found in Eqs.~(\ref{eq:CondTensor2nd}), (\ref{eq:CondTensor3rdVG}) and (\ref{eq:CondTensor3rd}) in the appendix.
%

To characterize the dependence of the optical response on the number of bands, $N_b$, we define a convergence measure that quantifies the difference with respect to the evaluation with a large number of bands. 
In our calculations, the reference number was set to $N_\mathrm{ref} = 20$. 
For the quantification, a \emph{truncation inaccuracy}, $\Delta(N_b)$, is defined as
\begin{align} 
	\label{eq:TruncationError}
	\Delta(N_b) & \equiv \dfrac{\expval{\big|\sigma(\omega,N_b)-\sigma(\omega,N_\mathrm{ref})\big|^2}}{ \expval{|\sigma(\omega,N_\mathrm{ref})|^2}} \, ,
\end{align}
where $\expval{|\sigma(\omega)|^2} \equiv \int_{\omega_{i}}^{\omega_{f}} |\sigma(\omega)|^2 \dd{\omega}/(\omega_{f}-\omega_{i})$ with $\omega_i$ and $\omega_f$ as the integration bounds.

\section{Results and Discussions \label{sec:Results}}
In this section, we address the dependence of the optical conductivity and several nonlinear processes on the basis truncation. 
Given the symmetry properties of the honeycomb lattice for hBN and restricting the external field to the in-plane directions (the crystal plane), it is sufficient to consider the diagonal components of the first-, second- \cite{Hipolito2016} and third-order \cite{Yang1995} conductivity tensors, namely $\sigma_{xx}^{(1)}$, $\sigma_{xxx}^{(2)}$, and $\sigma_{xxxx}^{(3)}$.
It should be noted that to ensure an adiabatic turn-on of the field, a positive infinitesimal value, $\eta = 0^+$, should be added to the frequency, \ie $\omega_p \rightarrow \omega_p+i\eta$. Throughout the paper, we set $\hbar\eta=\SI{0.03}{eV}$ to account for line broadening. Regarding the integration over the BZ, we discretize the rectangular area of Fig.~\ref{fig:Schematic}(b), which is equivalent to the first BZ, by at least 140000 $k$-points.
We start by presenting results for the linear response (optical conductivity), then proceed to the second-order interactions SHG and OR, and finally the third-order response. At the end, we compare quantitatively the truncation inaccuracy of the computed linear and nonlinear spectra.

\begin{figure}[t]
	\includegraphics[width=0.49\textwidth]{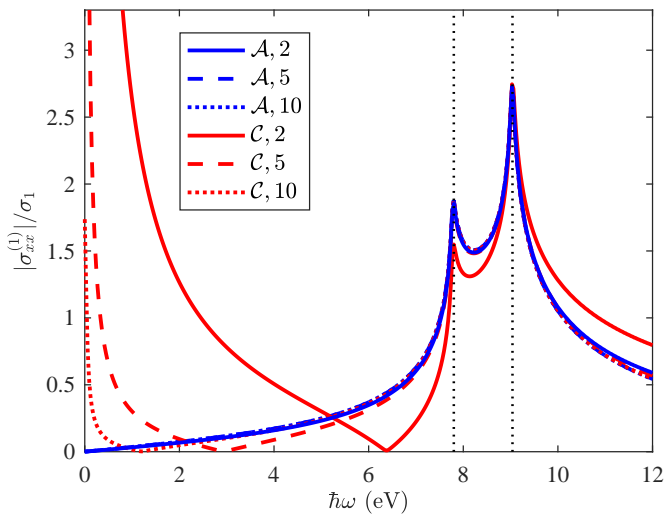}
	\caption{OC spectrum of hBN monolayer obtained from methods $\scase{A}$ (blue) and $\scase{C}$ (red) for $N_b=2$ (solid lines), $N_b=5$ (dashed lines), and $N_b=10$ (dotted lines). The values are normalized to $\sigma_1 \equiv e^2/4\hbar=\SI{6.0853e-5}{S}$. The black dotted lines indicate $\hbar\omega=\{E_g, E_\mathrm{vH}\}$.}
	\label{fig:Spectrum1st}
\end{figure}


\textbf{Linear response:} 
From the onset, Eqs.~(\ref{eq:FirstOrderA})-(\ref{eq:FirstOrderD}) show that in the presence of time-reversal symmetry the non-diagonal components vanish \cite{Hughes1996}. The results obtained from methods $\scase{D}$ and $\scase{A}$ are equivalent and neither introduces unphysical divergences in the evaluation of the current density response of cold insulators \cite{Aversa1995}.
In addition, $\sigma^{\scase{A}(1)}$ can be shown to be formally equivalent to $\sigma^{\scase{B}(1)}$ regardless of the basis size. The intraband parts of $\sigma^{\scase{A}(1)}$ and $\sigma^{\scase{B}(1)}$ are identical simply due to the well-known result $m \partial E_n/\partial k^\lambda = \hbar p_{nn}^\lambda$ \cite{Ashcroft1976}. The interband part of $\sigma_{\alpha\alpha}^{\scase{A}(1)}$ can be rewritten as
\begin{align}
	\label{eq:FirstEquiv}
	&C_e\sum_{\substack{\va{k},n,m \\ n \neq m}} \dfrac{p^\alpha_{nm} g_{mn}^\alpha}{E_{mn}} = \dfrac{C_e}{2} \bigg[ \sum_{\substack{\va{k},n,m \\ n \neq m}} \dfrac{p^\alpha_{nm} g_{mn}^\alpha}{E_{mn}} + (n \leftrightarrow m) \bigg]  \nonumber \\
	&=C_e\sum_{\substack{\va{k},n,m \\ n \neq m}} \dfrac{f_{nm}|p^\alpha_{mn}|^2}{E_{mn}} \dfrac{\hbar\omega_p}{(\hbar\omega_p)^2-E_{mn}^2}  \, ,
\end{align}
where in the first line, $(n \leftrightarrow m)$ indicates that the preceding term should be written with exchanged dummy indices $n$ and $m$. By the same token, the interband part of $\sigma_{\alpha\alpha}^{\scase{B}(1)}$ can be rewritten as Eq.~(\ref{eq:FirstEquiv}). 
With respect to the result derived with method $\scase{C}$, it can be shown to be equivalent to that of $\scase{A}$, if and only if a complete basis set is used \cite{Ventura2017}.

The truncation of the basis set breaks the equivalence between $\scase{A}$ and $\scase{C}$, leading to deviations between the optical response computed in these two methods. 
In Fig.~\ref{fig:Spectrum1st}, we compare the frequency dependence of the OC magnitude computed with methods $\scase{A}$ and $\scase{C}$ for three basis sets, $N_b = \{2,5,10\}$.
%
The variation of the length gauge results is quite small and not visible on the scale of Fig.~\ref{fig:Spectrum1st}.
In contrast, the velocity gauge response, $\sigma^{\scase{C}(1)}$, is strongly dependent on the number of bands. In particular, the zero-frequency divergence is strongly suppressed with increasing $N_b$. Notwithstanding this strong suppression, the zero-frequency divergence remains present for any finite basis set. 
In addition, the features associated with the band gap and van Hove  (vH) singularity also converge to the results computed using the length gauge.

\begin{figure}[t]
	\includegraphics[width=0.49\textwidth]{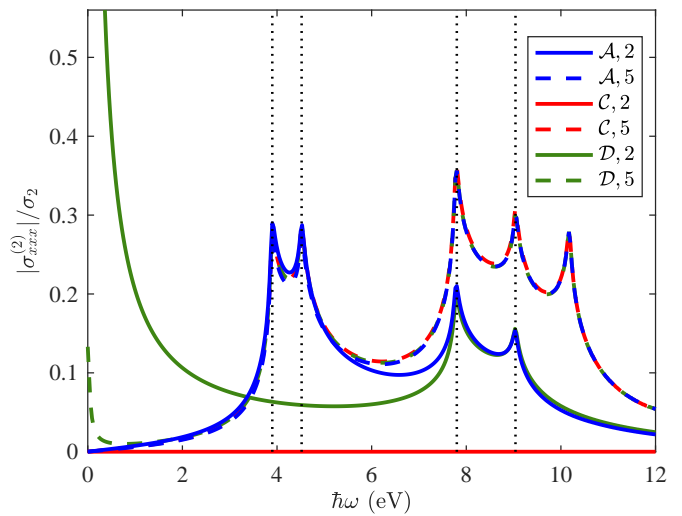}
	\caption{SHG spectrum of hBN monolayer obtained from methods $\scase{A}$ (blue), $\scase{C}$ (red) and $\scase{D}$ (green), for $N_b=2$ (solid lines) and $N_b=5$ (dashed lines).
		The values are normalized to $\sigma_2 \equiv e^3a/8\gamma_0\hbar=\SI{3.2797e-15}{SmV^{-1}}$. The black dotted lines from left to right mark $2\hbar\omega=E_g$, $2\hbar\omega=E_\mathrm{vH}$, $\hbar\omega=E_g$, and $\hbar\omega =E_\mathrm{vH}$, respectively.}
	\label{fig:Spectrum2nd}
\end{figure}

\textbf{Second-order response:} 
The conductivity tensors for the four methods are shown in Eq.~(\ref{eq:CondTensor2nd}) (see Appendix~\ref{sec:AppendixA}). Based on these expressions, we numerically demonstrate the equivalence of Eqs.~(\ref{eq:CondTensor2ndA})-(\ref{eq:CondTensor2ndD}) for a large basis set.  
Figure~\ref{fig:Spectrum2nd} illustrates the SHG conductivities, $\sigma_{xxx}^{(2)}(\omega,\omega)$, for two representative sizes of the basis set, namely $N_b = \{2,5\}$. The results obtained by method $\scase{B}$ are numerically identical to method $\scase{A}$ and, hence, are omitted from the figure.
Considering the response in the vicinity of the lower energy features, \ie $2\hbar\omega \lesssim E_\mathrm{vH}$, our results show that the calculation based on method $\scase{A}$ exhibits a small dependence on the basis size.
In striking contrast, both methods $\scase{C}$ and $\scase{D}$ present highly different results. The former is identically zero for all frequencies due to time-reversal symmetry. Regarding the latter, it is non-zero but exhibits a zero-frequency divergence and does not reproduce the SHG features associated with $2\hbar\omega \sim \{E_g, E_\mathrm{vH}\}$.
At higher energies, $\hbar\omega \sim \{E_g,E_\mathrm{vH}\}$, the responses computed with the four methods show strong variations with the increase of $N_b$. 
It is important to note that this variation arises from interactions between the valence band and the second conduction band, \ie $2\hbar\omega \sim E_{31}$ (see Fig.~\ref{fig:BandStructure}). Hence, the deviation of the spectra in this frequency range should not be considered a convergence issue. Rather, they are a consequence of the limited frequency range, for which the two-band model is applicable.

\begin{figure}[t]
	\includegraphics[width=0.49\textwidth]{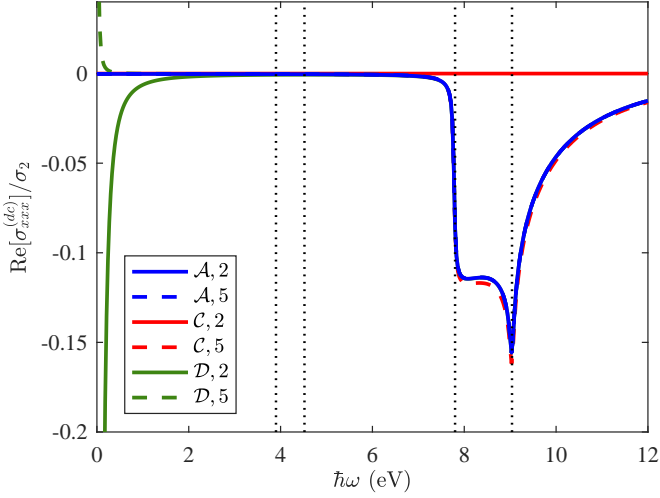}
	\caption{OR spectrum of hBN monolayer obtained from methods $\scase{A}$ (blue), $\scase{C}$ (red) and $\scase{D}$ (green), for $N_b=2$ (solid lines) and $N_b=5$ (dashed lines). The values are normalized to $\sigma_2$ (see Fig.~\ref{fig:Spectrum2nd} caption). The dotted lines from the left to right indicate $2\hbar\omega=E_g$, $2\hbar\omega=E_\mathrm{vH}$, $\hbar\omega=E_g$, and  $\hbar\omega=E_\mathrm{vH}$, respectively.}
	\label{fig:SpectrumDC}
\end{figure}

The basis truncation also affects the calculation of other second-order processes such as OR, $\sigma_{xxx}^{(2)}(\omega,-\omega)$, as illustrated in Fig.~\ref{fig:SpectrumDC}.
The OR results are similar to those obtained for SHG, but given the fact that the OR conductivity does not contain a $2\hbar\omega-E_{cv}$ term in denominator, the response starts at $\hbar\omega \sim E_g$. 
In this case, the most significant differences appear in method $\scase{C}$ where, similarly to the SHG process, the two-band calculation yields a zero response regardless of the external photon frequency. Moreover, the results generated by method $\scase{D}$ feature once more a spurious zero-frequency divergence, which is suppressed gradually by including more bands in the calculations.

For the results in Figs.~\ref{fig:Spectrum2nd} and \ref{fig:SpectrumDC}, we evaluate the GDs appearing in Eqs.~(\ref{eq:CondTensor2nd}) by employing the definition, Eq.~(\ref{eq:GenDerivative}). However, as pointed out in section~\ref{sec:OpticalResponse}, one may employ the sum rule, Eq.~(\ref{eq:SumRule}), for evaluating the GDs presented in the intraband parts of $\sigma^{\scase{A}(2)}$, $\sigma^{\scase{B}(2)}$, and $\sigma^{\scase{D}(2)}$. Following this substitution, up to the machine precision, all the three methods generate SHG and OR spectra identical to those of method $\scase{C}$ for any size of basis set. This means that, for a two-band model, the length gauge approach produces identically zero response similar to the velocity gauge if one uses the sum rule for evaluating the GD. Thus, the choice between exact and approximate implementation of the GD is of considerable importance for a truncated basis set.

\begin{figure}[t]
	\includegraphics[width=0.49\textwidth]{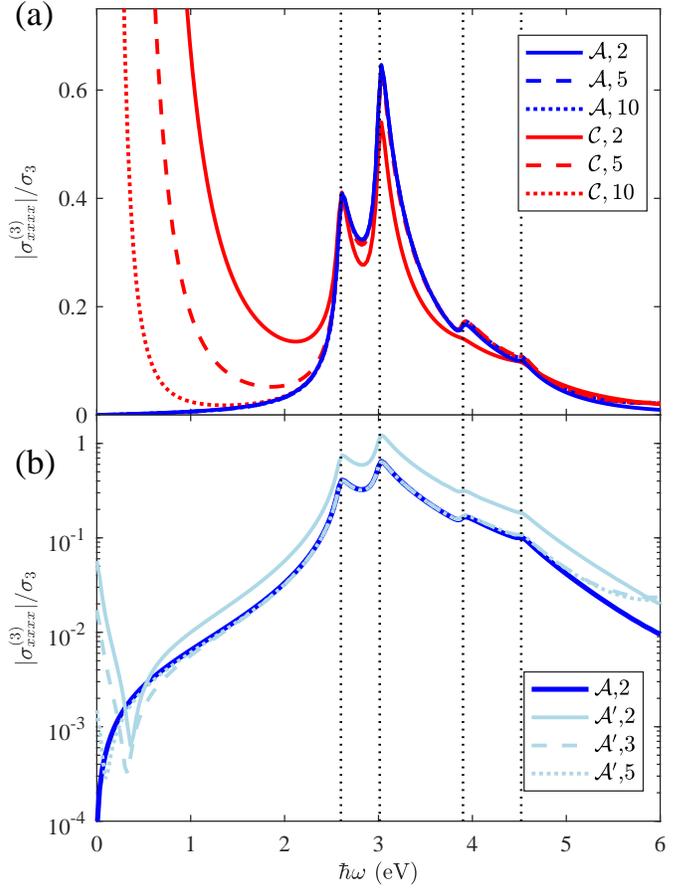}
	\caption{(a) THG spectrum of hBN monolayer obtained from methods $\scase{A}$ (blue) and $\scase{C}$ (red) for basis sets with size of $N_b=2$ (solid lines), $N_b=5$ (dashed lines), and $N_b=10$ (dotted lines). The values are normalized to $\sigma_3 \equiv e^4a^2/16\gamma_0^2\hbar= \SI{1.7675e-25}{Sm^2V^{-2}}$. The black dotted lines from the left mark $3\hbar\omega=E_g$, $3\hbar\omega=E_\mathrm{vH}$, $2\hbar\omega=E_g$, and $2\hbar\omega=E_\mathrm{vH}$, respectively. (b) THG spectrum of hBN monolayer (on log-scale) obtained from method $\scase{A}$ by the direct evaluation of GD, Eq.~(\ref{eq:GenDerivative}), (blue solid line) for $N_b=2$ or by employing the sum rule, Eq.~(\ref{eq:SumRule}), (light blue), labeled as $\scase{A}'$, for $N_b=2$ (solid line), $N_b=3$ (dashed line), and $N_b=5$ (dotted line).}
	\label{fig:Spectrum3rdAll}
\end{figure}

\begin{figure*}[t]
	\includegraphics[width=\textwidth]{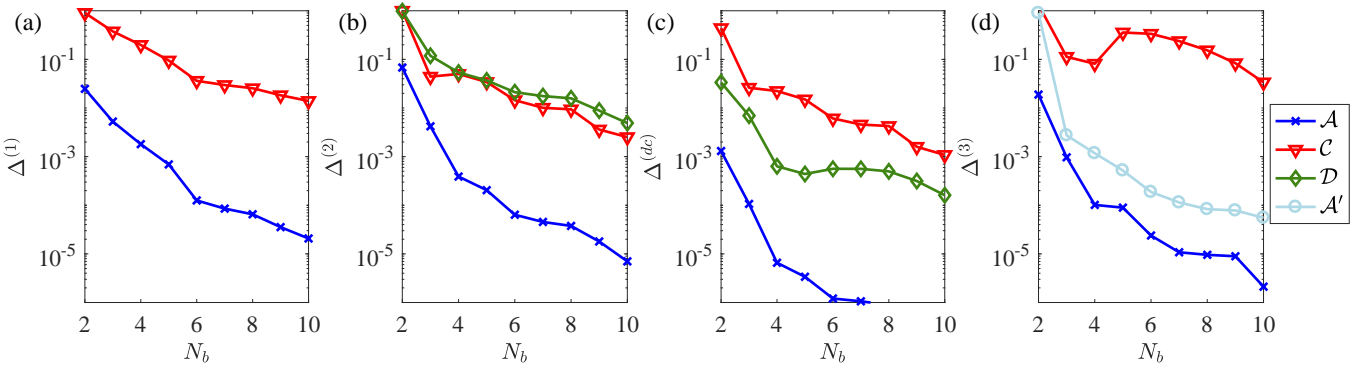}
	\caption{Truncation inaccuracy, $\Delta(N_b)$, defined in Eq.~(\ref{eq:TruncationError}), for the four processes under consideration, namely (a) OC, (b) SHG, (c) OR, and (d) THG, obtained by employing methods $\scase{A}$ (blue), $\scase{C}$ (red), and $\scase{D}$ (green). Method $\scase{A}'$ (light blue) indicates the modified $\scase{A}$, for which the GDs are approximated by Eq.~(\ref{eq:SumRule}). Reference results are computed with $N_{ref} = 20$. The lower spectral limit for the analysis is set at $\hbar\omega_i=0.1E_g$ and the upper limits read: (a) $\hbar\omega_f=1.5E_g$, (b) $\hbar\omega_f=0.7E_g$, (c) $\hbar\omega_f=1.5E_g$, (d) $\hbar\omega_f=0.7E_g$.}
	\label{fig:Convergence}
\end{figure*}

\textbf{Third-order response:} 
Here, we limit our analysis to the effects of truncation in the calculation of THG, $\sigma_{xxxx}^{(3)}(\omega,\omega,\omega)$, via methods $\scase{A}$ and $\scase{C}$.
In Fig.~\ref{fig:Spectrum3rdAll}(a), we compare the THG conductivity computed by both methods for three values of $N_b$. The THG results follow the trends observed in the linear and quadratic responses. 
Firstly, the results computed via method $\scase{A}$ display only a weak dependence on the size of the truncated basis. Secondly, the velocity gauge results exhibit strong zero-frequency divergences, that reduce, although slowly, with the increasing number of bands. Compared to the linear response, the divergence at zero frequency of method $\scase{C}$ in the third-order spectrum is stronger, \ie as $\omega^{-3}$ vs. $\omega^{-1}$. 

In Fig.~\ref{fig:Spectrum3rdAll}(b), we compare the THG spectra obtained with method $\scase{A}$ either by calculating the GDs directly using Eq.~(\ref{eq:GenDerivative}) or by employing the sum rule Eq.~(\ref{eq:SumRule}), labeling the latter approach as $\scase{A}'$. The spectra are plotted on a log-scale to be more illustrative. For $N_b=2$, the results show that the approximate GD from Eq.~(\ref{eq:SumRule}) overestimates the response in the frequency range of $3\hbar\omega \sim \{E_g, E_\mathrm{vH}\}$ and $2\hbar\omega \sim \{E_g, E_\mathrm{vH}\}$. By increasing $N_b$ to 3 and then 5, however, the 2-5 eV features of method $\scase{A}'$ converge to those of method $\scase{A}$. Even so, the low-frequency response in $\scase{A}'$ still deviates considerably from method $\scase{A}$, which demonstrates the need for a large basis set. As in the SHG case, it should be noted that the high-frequency ($\hbar \omega > 5$ eV) deviation can be attributed to transitions involving higher conduction bands and, hence, is of less importance. 

\textbf{Convergence analysis:} 
Figures~\ref{fig:Convergence}(a) to \ref{fig:Convergence}(d) show the truncation inaccuracy $\Delta(N_b)$ defined in Eq.~(\ref{eq:TruncationError}), as a function of basis size $N_b$ for linear, SHG, OR, and THG response of monolayer hBN, respectively. 
Since the two length gauge approaches $\scase{A}$ and $\scase{B}$ generate numerically identical spectra for all linear and nonlinear processes, the truncation inaccuracy of method $\scase{B}$ is omitted. Note, however, that method $\scase{B}$ typically requires additional efforts compared to $\scase{A}$ in its numerical implementation due to the additional position operator. The convergence behavior of all four investigated linear and nonlinear processes is qualitatively similar. For instance, the length gauge approaches with the direct evaluation of GDs converge faster with respect to the basis set size than the velocity gauge methods, \ie $\scase{C}$ and $\scase{D}$, in all cases. If the sum rule of Eq.~(\ref{eq:SumRule}) is employed for evaluating the GD appearing in methods $\scase{A}$, $\scase{B}$ and $\scase{D}$, the truncation inaccuracies will be identical to that of the method $\scase{C}$ for the second-order responses.  
Moreover, the truncation inaccuracy computed in the velocity gauge, method $\scase{C}$, increases significantly for the third-order response due to the strong zero-frequency divergence. Also, for the third-order responses, the modified $\scase{A}$ method, $\scase{A}'$, generates more accurate spectra than method $\scase{C}$, when at least 3 bands are used in the calculation. Nevertheless, it underperforms when compared with the original method $\scase{A}$ for all basis sizes considered.

\section{Summary \label{sec:Conclusion}}
In summary, we have investigated the effects of basis truncation on several linear and nonlinear optical response functions including the OC, SHG, OR and THG. The conductivity tensors are derived and compared using four computationally different approaches. These result from combining two choices of gauges and two ways of evaluating the current density, \ie directly or via the polarization. For the OC, the equivalence of all four methods has been demonstrated analytically provided a complete basis set is used, whereas for the NLO response, we have demonstrated it numerically by employing a large basis set in calculations. The length gauge approaches, \ie tensors labeled with $\scase{A}$ or $\scase{B}$, generate the most accurate spectra, compared to the velocity gauge approaches, particularly for small basis sets. In addition, it has been shown that the choice of method to compute the GD is crucial as the evaluation based on the sum rule, Eq.~(\ref{eq:SumRule}), may result in degrading convergence. Finally, although, the well-known zero-frequency divergences in the velocity gauge responses vanish by increasing the size of basis set, the calculated spectra are far less accurate than the ones generated by the length gauge methods. Our results shed light on the source of the differences arising from several commonly used computational approaches to the linear and nonlinear optical response.

\acknowledgments
The authors thank F. Bonabi for helpful discussions throughout the project. This work was supported by the QUSCOPE center sponsored by the Villum Foundation and TGP is financially supported by the CNG center under the Danish National Research Foundation, project DNRF103.

\appendix

\begin{widetext}
\section{Nonlinear conductivity tensors \label{sec:AppendixA}}
\subsection{Perturbative density matrix \label{sec:DensityMatrix}}
The equation of motion can be solved by employing a perturbative approach and expanding the solution as a power series of the perturbation: $\rho_{mn}(t) = \sum_{N=0}^{\infty} \rho_{mn}^{(N)}(t)$, where $\rho_{mn}^{(N)}(t)$ at order $N$ is determined iteratively from the previous order via
\begin{equation}
	\label{eq:Pertubation}
	\rho_{mn}^{(N)}(t) = \dfrac{1}{i\hbar} \int_{-\infty}^{t} \comm{\hat{V}(t)}{\hat{\rho}^{(N-1)}(t)}_{mn} e^{iE_{mn}(t'-t)/\hbar} \dd{t'} \, .
\end{equation}
In the absence of any perturbation, the system is at equilibrium and its density matrix elements are determined by $\rho_{mn}^{(0)} \equiv \delta_{mn}f_n$.
Hence, by integrating Eq.~(\ref{eq:Pertubation}) for a set of time harmonic perturbations, $\hat{V}(t) = 1/2\sum_p \hat{V}(\omega_p) \exp(-i\omega_p t)$, the first three terms read \cite{Boyd2008}
\begin{subequations}
	\label{eq:DensityElements}
	\begin{align}
		\label{eq:DensityA} \rho_{mn}^{(1)}(t) &= \dfrac{1}{2} \sum_p f_{nm} \frac{ V_{mn}( \omega_p ) }{ \hbar\omega_p -E_{mn} } e^{ -i\omega_p t }, \\
		\label{eq:DensityB} \rho_{mn}^{(2)}(t) &= \dfrac{1}{4} \sum_{p,q} \sum_l \dfrac{ e^{ -i( \omega_p +\omega_q )t } }{\hbar( \omega_p +\omega_q ) -E_{mn}} \bigg[ \dfrac{ f_{nl} V_{ml}( \omega_q ) V_{ln} ( \omega_p ) }{\hbar\omega_p -E_{ln} } -\dfrac{ f_{lm} V_{ml}( \omega_p ) V_{ln} ( \omega_q ) }{\hbar\omega_p -E_{ml} } \bigg], \\ 
		\label{eq:DensityC} \rho_{mn}^{(3)}(t) &=\dfrac{1}{8} \sum_{p,q,s} \sum_{j,l} \dfrac{ e^{ -i( \omega_p +\omega_q +\omega_s )t } }{\hbar( \omega_p +\omega_q +\omega_s ) -E_{mn} }\Bigg\{ \dfrac{ V_{mj} ( \omega_p ) }{ \hbar( \omega_q +\omega_s ) -E_{jn} } \bigg[ \dfrac{ f_{nl} V_{jl}( \omega_q ) V_{ln} ( \omega_s ) }{ \hbar\omega_s -E_{ln} } -\dfrac{ f_{lj} V_{jl}( \omega_s ) V_{ln} ( \omega_q ) }{\hbar\omega_s -E_{jl} } \bigg] \nonumber \\ 
		& +\dfrac{ V_{ln}( \omega_p ) }{ \hbar( \omega_q +\omega_s ) -E_{ml} } \bigg[ \dfrac{ f_{jm} V_{mj}( \omega_s ) V_{jl}( \omega_q ) }{\hbar\omega_s -E_{mj} } - \dfrac{ f_{lj} V_{mj}( \omega_q ) V_{jl}( \omega_s ) }{\hbar\omega_s -E_{jl} } \bigg] \Bigg\},  
	\end{align}
\end{subequations}
where $V_{mn}$ are the matrix elements of the perturbation, \ie  $V_{mn}(\omega) \equiv \mel{m}{\hat{V}(\omega)}{n}$. Two choices of gauge are employed for the perturbative Hamiltonian $\hat{V}$, \ie $\hat{V}_l$ and $\hat{V}_v$ as defined in section~\ref{sec:OpticalResponse}. 

\subsection{Second-order tensors}
Here, we show the tensor expressions derived for the four methods of Table~\ref{tb:Gauges} using Eq.~(\ref{eq:DensityB}). The expression for method $\scase{C}$ is obtained straightforwardly, since it only contains the matrix elements of the well-defined momentum operator. For the methods $\scase{A}$, $\scase{B}$ and $\scase{D}$, the position operator has to be separated into its interband/intraband parts, and it should be treated carefully as outlined briefly in section~\ref{sec:OpticalResponse}, and in detail in Ref.~\onlinecite{Aversa1995}. 
\begin{subequations}
	\label{eq:CondTensor2nd}
	\begin{align}
	\label{eq:CondTensor2ndA} \sigma_{\lambda\alpha\beta}^{\scase{A}(2)}(\omega_p,\omega_q) &\equiv C_{ee} \sum_{\substack{\va{k},n,m,l \\ n \neq l \neq m}} \dfrac{p_{nm}^\lambda \big( g_{ln}^\alpha p_{ml}^\beta - g_{ml}^\alpha p_{ln}^\beta \big)}{E_{ml} E_{ln} [\hbar(\omega_p+\omega_q)-E_{mn}]} 
	+ C_{ie} \sum_{\substack{\va{k},n,m \\ n \neq m}} \dfrac{-p_{nm}^\lambda}{\hbar(\omega_p+\omega_q)-E_{mn}} \bigg( \dfrac{g_{mn}^{\alpha}}{E_{mn}} \bigg)_{;k^\beta} \nonumber \\
	&+ C_{ie} \sum_{\substack{\va{k},n,m \\ n \neq m}} \dfrac{-p_{nm}^\lambda}{E_{mn}[\hbar(\omega_p+\omega_q)-E_{mn}]} \dfrac{p_{mn}^\beta}{\hbar\omega_p} \pdv{f_{nm}}{k^\alpha} 
	+ C_{ii}\sum_{\va{k},n} \dfrac{p_{nn}^\lambda}{\hbar(\omega_p+\omega_q)(\hbar\omega_p)} \pdv{f_n}{k^\beta}{k^\alpha} \, , \\
	\label{eq:CondTensor2ndB} \sigma_{\lambda\alpha\beta}^{\scase{B}(2)}(\omega_p,\omega_q) &\equiv C_{ee} \sum_{\substack{\va{k},n,m,l \\ n \neq m \neq l \neq n}} \dfrac{\hbar(\omega_p+\omega_q)p_{nm}^\lambda \big( g_{ln}^\alpha p_{ml}^\beta - g_{ml}^\alpha p_{ln}^\beta \big)}{E_{mn} E_{ml} E_{ln} [\hbar(\omega_p+\omega_q)-E_{mn}]} 
	+ C_{ie} \sum_{\substack{\va{k},n,m \\ n \neq m}}  \dfrac{-\hbar(\omega_p+\omega_q)p_{nm}^\lambda}{E_{mn}[\hbar(\omega_p+\omega_q)-E_{mn}]} \bigg( \dfrac{g_{mn}^{\alpha}}{E_{mn}} \bigg)_{;k^\beta} \nonumber \\
	&+C_{ie} \sum_{\substack{\va{k},n,m \\ n \neq m}} \dfrac{-\hbar(\omega_p+\omega_q)}{2E_{mn}^2} \dfrac{g_{mn}^\alpha (p_{nm}^\beta)_{;k^\lambda}}{\hbar\omega_q+E_{mn}} 
	+ C_{ie} \sum_{\substack{\va{k},n,m \\ n \neq m}} \dfrac{-\hbar(\omega_p+\omega_q)p_{nm}^\lambda}{E_{mn}^2[\hbar(\omega_p+\omega_q)-E_{mn}]} \dfrac{p_{mn}^\beta}{\hbar\omega_p} \pdv{f_{nm}}{k^\alpha} \, , \\
	\label{eq:CondTensor2ndC} \sigma_{\lambda\alpha\beta}^{\scase{C}(2)}(\omega_p,\omega_q) &\equiv C_{ee}\dfrac{1}{(\hbar\omega_p) (\hbar\omega_q)} \sum_{\va{k},n,m,l} \dfrac{p_{nm}^\lambda}{\hbar(\omega_p+\omega_q)-E_{mn}} \big( g_{ln}^\alpha 
	p_{ml}^\beta - g_{ml}^\alpha p_{ln}^\beta \big) \, , \\
	\label{eq:CondTensor2ndD} \sigma_{\lambda\alpha\beta}^{\scase{D}(2)}(\omega_p,\omega_q) &\equiv C_{ee}\dfrac{\hbar(\omega_p+\omega_q)}{(\hbar\omega_p) (\hbar\omega_q)} \sum_{\substack{\va{k},n,m,l \\ n \neq m}} \dfrac{p_{nm}^\lambda \big( g_{ln}^\alpha p_{ml}^\beta - g_{ml}^\alpha p_{ln}^\beta \big)}{\hbar(\omega_p+\omega_q)-E_{mn}} 
	+ C_{ie} \dfrac{\hbar(\omega_p+\omega_q)}{2(\hbar\omega_p)(\hbar\omega_q)} \sum_{\va{k},n,m} \dfrac{g_{mn}^\alpha (p_{nm}^\beta)_{;k^\lambda}}{\hbar\omega_q+E_{mn}} \, ,
	\end{align}
\end{subequations}
where $g_{mn}^\alpha$ has been defined in Eq.~(\ref{eq:FirstConstants}) and the constants $C_{ee}$, $C_{ie}$ and $C_{ii}$ read
\begin{align}
	C_{ee} \equiv C_{ie} \dfrac{\hbar}{ m } \equiv C_{ii} \dfrac{\hbar^2}{ m ^2} \equiv \dfrac{g e ^3 \hbar^2}{4 m ^3 A} \, .
\end{align}  
The expression for tensor $\sigma_{\lambda\alpha\beta}^{\scase{A}(2)}$ consists of four 
terms: one purely-interband, two mixed interband-intraband, and one purely-intraband contribution, respectively.
By the same token, similar interband and intraband contributions in $\sigma_{\lambda\alpha\beta}^{\scase{B}(2)}$ can be identified. 
In contrast, the interband/intraband contributions are not separated in the expression for $\sigma_{\lambda\alpha\beta}^{\scase{C}(2)}$, and they are only partly divided in $\sigma_{\lambda\alpha\beta}^{\scase{D}(2)}$ due to the presence of $\va{\hat{r}}$ in the current density operator.
For a cold, intrinsic semiconductor, the terms including derivatives of the band population $f_n$ vanish, \eg the last two terms of $\sigma_{\lambda\alpha\beta}^{\scase{A}(2)}$ and the last part of $\sigma_{\lambda\alpha\beta}^{\scase{B}(2)}$. The conductivity expressions in Eqs.~(\ref{eq:CondTensor2nd}) can be symmetrized with respect to the frequencies and indices by performing a permutation \cite{Sipe2000}.
In deriving the conductivity tensors, the following useful expressions that can be derived from Eq.~(\ref{eq:GenDerivative}) have been used:
\begin{subequations}
	\begin{align}
		&\lim_{\va{k} \rightarrow \va{k}'} \big(f_{n\va{k}}-f_{m\va{k}'} \big) \va{r}_{mn}^{(i)} = i \delta_{nm} \gradient_{\va{k}} f_{n} \, ,\\
		&\big[ \va{r}_{nn}^{(i)}-\va{r}_{mm}^{(i)} \big] p_{nm}^\alpha = i \delta_{\va{k}\va{k}'} \big(p_{nm}^\alpha\big)_{;\va{k}} \, .
	\end{align}
\end{subequations}
Finally, the GDs appearing in Eqs.~(\ref{eq:CondTensor2ndA}), (\ref{eq:CondTensor2ndB}) and (\ref{eq:CondTensor2ndD}) can be computed either directly from the definition Eq.~(\ref{eq:GenDerivative}) or by using the sum rule Eq.~(\ref{eq:SumRule}). For the latter, the chain rule property of the GD and an additional expression $(E_{mn})_{;k^\alpha} = \hbar(p_{mm}^\alpha-p_{nn}^\alpha)/m$ are utilized \cite{Aversa1995}.

\subsection{Third-order tensors}
The third-order conductivity tensor components $\sigma_{\lambda\alpha\beta\gamma}^{(3)}$ are derived by inserting the length and velocity gauge perturbative Hamiltonian into the density matrix elements of Eq.~(\ref{eq:DensityC}). 
Here, we only present the conductivity tensors obtained in methods $\scase{A}$ and $\scase{C}$. As in the case of the second-order tensor, $\sigma_{\lambda\alpha\beta\gamma}^{\scase{C}(3)}$ is obtained straightforwardly as
\begin{equation}
	\label{eq:CondTensor3rdVG} \sigma_{\lambda\alpha\beta\gamma}^{\scase{C}(3)}(\omega_p,\omega_q,\omega_s) \equiv C_{eee} \dfrac{1}{(\hbar\omega_p)(\hbar\omega_q)(\hbar\omega_s)} \sum_{\va{k},n,m,j,l} \dfrac{p_{nm}^\lambda}{\hbar\omega_3-E_{mn}} \Bigg[ \dfrac{p_{mj}^\alpha \big( g_{ln}^\gamma p_{jl}^\beta - g_{jl}^\gamma p_{ln}^\beta \big)}{\hbar\omega_2-E_{jn}}  
	+ \dfrac{p_{ln}^\alpha\big( g_{mj}^\gamma p_{jl}^\beta - g_{jl}^\gamma p_{mj}^\beta, \big)}{\hbar\omega_2-E_{ml}} \Bigg] \, ,
\end{equation}
where we introduce the coefficient $C_{eee} \equiv g e^4\hbar^3/(i8m^4 A)$ and auxiliary variables $\omega_2 \equiv \omega_q+\omega_s$ $\omega_3 \equiv \omega_p+\omega_q+\omega_s$, $g_{mn}^{\beta,\gamma} \equiv f_{nm} p^{\beta,\gamma}_{mn}/(\hbar\omega_s-E_{mn})$.
On the other hand, the calculation of $\sigma_{\lambda\alpha\beta\gamma}^{\scase{A}(3)}$ requires dividing the ill-defined position operator into interband and intraband parts. The resulting eight different combinations of interband/intraband terms, denoted by $eee$, $eei$, $eie$, $eii$, $iee$, $iei$, $iie$, $iii$, are given by
\begin{subequations}
	\label{eq:CondTensor3rd}
	\begin{align}
		\label{eq:CondTensor3rdA} &\sigma_{\lambda\alpha\beta\gamma}^{\scase{A}(3)}(\omega_p,\omega_q,\omega_s) \equiv \sigma_{\lambda\alpha\beta\gamma}^{(3,eee)}+\sigma_{\lambda\alpha\beta\gamma}^{(3,eei)}+\sigma_{\lambda\alpha\beta\gamma}^{(3,eie)}+\sigma_{\lambda\alpha\beta\gamma}^{(3,eii)}+\sigma_{\lambda\alpha\beta\gamma}^{(3,iee)}+\sigma_{\lambda\alpha\beta\gamma}^{(3,iei)}+\sigma_{\lambda\alpha\beta\gamma}^{(3,iie)}+\sigma_{\lambda\alpha\beta\gamma}^{(3,iii)} \, , \\ 
		\label{eq:CondTensor3rdB} &\sigma_{\lambda\alpha\beta\gamma}^{(3,eee)} \equiv C_{eee} \sum_{\substack{\va{k},n,m,j,l \\ n \neq l \neq j \neq m}} \dfrac{p_{nm}^\lambda}{\hbar\omega_3-E_{mn}} \Bigg[ \dfrac{p_{mj}^\alpha \big( g_{ln}^\gamma p_{jl}^\beta - g_{jl}^\gamma p_{ln}^\beta \big)}{E_{mj}E_{jl}E_{ln}(\hbar\omega_2-E_{jn})} 
		+ \dfrac{p_{ln}^\alpha\big( g_{mj}^\gamma p_{jl}^\beta - g_{jl}^\gamma p_{mj}^\beta \big)}{E_{mj}E_{jl}E_{ln}(\hbar\omega_2-E_{ml})} \Bigg] \, , \\ 
		\label{eq:CondTensor3rdC} &\sigma_{\lambda\alpha\beta\gamma}^{(3,eei)} \equiv C_{iee} \dfrac{1}{\hbar\omega_s} \sum_{\substack{\va{k},n,m,l \\ n \neq l \neq m}} \dfrac{-p_{nm}^\lambda}{E_{ln}E_{ml} (\hbar\omega_3 -E_{mn})} \Bigg[ \dfrac{ p_{ml}^\alpha p_{ln}^\beta}{\hbar\omega_2-E_{ln}} \pdv{f_{nl}}{k^\gamma}   
		+\dfrac{p_{ln}^\alpha p_{ml}^\beta}{\hbar\omega_2-E_{ml}} \pdv{f_{ml}}{k^\gamma} \Bigg] \, , \\
		\label{eq:CondTensor3rdD} &\sigma_{\lambda\alpha\beta\gamma}^{(3,eie)} \equiv C_{iee} \sum_{\substack{\va{k},n,m,l \\ n \neq l \neq m}} \dfrac{p_{nm}^\lambda}{\hbar\omega_3-E_{mn}} \Bigg[ \dfrac{ p_{ml}^\alpha/E_{ml}}{\hbar\omega_2-E_{ln}} \bigg( \dfrac{g_{ln}^\gamma }{E_{ln}} \bigg)_{;k^\beta} 
		- \dfrac{ p_{ln}^\alpha/E_{ln}}{\hbar\omega_2-E_{ml}} \bigg( \dfrac{g_{ml}^\gamma}{E_{ml}} \bigg)_{;k^\beta} \Bigg] \, , \\ 
		\label{eq:CondTensor3rdE} &\sigma_{\lambda\alpha\beta\gamma}^{(3,eii)} \equiv C_{iie} \dfrac{1}{(\hbar\omega_s)(\hbar\omega_2)} \sum_{\substack{\va{k},n,m \\ n \neq m}} \dfrac{p_{nm}^\lambda p_{mn}^\alpha/E_{mn}}{(\hbar\omega_3-E_{mn})} \pdv{f_{nm}}{k^\beta}{k^\gamma} \, , \\
		\label{eq:CondTensor3rdF} &\sigma_{\lambda\alpha\beta\gamma}^{(3,iee)} \equiv C_{iee} \sum_{\substack{\va{k},n,m,l \\ n \neq l \neq m}} \Bigg[ \dfrac{p_{nm}^\lambda}{\hbar\omega_3-E_{mn}} \Bigg]_{;k^\alpha} \dfrac{ g_{ln}^\gamma p_{ml}^\beta - g_{ml}^\gamma p_{ln}^\beta}{E_{ml}E_{nl}(\hbar\omega_2-E_{mn})} \, , \\ 
		\label{eq:CondTensor3rdG} &\sigma_{\lambda\alpha\beta\gamma}^{(3,iei)} \equiv C_{iie} \dfrac{1}{\hbar\omega_s} \sum_{\substack{\va{k},n,m \\ n \neq m}} \dfrac{-p_{nm}^\lambda}{\hbar\omega_3-E_{mn}} \Bigg[ \dfrac{p_{mn}^\beta/E_{mn}}{\hbar\omega_2-E_{mn}} \pdv{f_{mn}}{k^\gamma}\Bigg]_{;k^\alpha} \, , \\
		\label{eq:CondTensor3rdH} &\sigma_{\lambda\alpha\beta\gamma}^{(3,iie)} \equiv C_{iie} \sum_{\substack{\va{k},n,m \\ n \neq m}} \bigg[\dfrac{p_{nm}^\lambda}{\hbar\omega_3-E_{mn}} \bigg]_{;k^\alpha} \dfrac{-1}{\hbar\omega_2-E_{mn}} \bigg( \dfrac{g_{mn}^\gamma}{E_{mn}} \bigg)_{;k^\beta}\, , \\
		\label{eq:CondTensor3rdI} &\sigma_{\lambda\alpha\beta\gamma}^{(3,iii)} \equiv C_{iii} \dfrac{1}{(\hbar\omega_s)(\hbar\omega_2)(\hbar\omega_3)} \sum_{\va{k},n} (-p_{nn}^\lambda) \dfrac{\partial^3 f_n}{\partial k^\alpha \partial k^\beta \partial k^\gamma} \, , 
	\end{align}
\end{subequations}
where $C_{iee} \hbar/m \equiv C_{iie} \hbar^2/m^2 \equiv C_{iii} \hbar^3/m^3 \equiv C_{eee}$ and integration by parts has been used in the $k-$summation when deriving Eqs.~(\ref{eq:CondTensor3rdF}) and ~(\ref{eq:CondTensor3rdH}). It should be noted that Eqs.~(\ref{eq:CondTensor3rdVG}) and (\ref{eq:CondTensor3rd}) have not been symmetrized with respect to the frequencies and indices, which can be performed by permutation of frequencies and component indices \cite{Hipolito2017b}. For a clean, cold semiconductor only the $eee$, $eie$, $iee$, and $iie$ terms are nonzero.
\end{widetext}

\bibliography{BN_Nonlinear}

\end{document}